\documentclass[10pt,conference]{IEEEtran}\IEEEoverridecommandlockouts
\usepackage{epsfig,graphics,subfigure,psfrag,amsmath,cases}
\usepackage{latexsym,amssymb,amsmath,epsfig,subfigure,algorithm}
\usepackage{algorithmic}
\usepackage{color}
\usepackage{url}
\usepackage{scrtime}

\author{\authorblockN{ Derrick Wing Kwan Ng\authorrefmark{1} and Robert Schober\authorrefmark{1}\thanks{This work was supported in part by the AvH Professorship Program of the Alexander von Humboldt Foundation.}}
\authorrefmark{1}Institute for Digital Communications, Universit\"at Erlangen-N\"urnberg, Germany\vspace*{-4mm}
}

\title{Spectral Efficient Optimization in OFDM Systems with Wireless  Information and Power Transfer }

\date{\thistime,\,\today}

\newcommand{\abs}[1]{\lvert#1\rvert}

\begin{document}

\maketitle

\begin{abstract}
This paper considers an orthogonal frequency division
multiplexing (OFDM) point-to-point wireless communication system with simultaneous wireless
information and power transfer. We study a receiver which is able
to harvest energy  from the desired signal, noise, and interference.  In particular, we consider a
power splitting receiver which dynamically splits the received power into two power streams for
information decoding and energy harvesting. We  design  power
allocation  algorithms maximizing the spectral
efficiency (bit/s/Hz) of data transmission.
In particular, the algorithm design is formulated as a nonconvex
optimization problem which takes into account the constraint on the
minimum power delivered to the receiver. The problem is
solved by using convex optimization techniques and a  one-dimensional search. The optimal power allocation algorithm serves as a system
benchmark scheme due to its high complexity.  To strike a balance
between system performance and computational complexity, we also propose two
suboptimal algorithms which require a low computational complexity.  Simulation results demonstrate the excellent
performance of the proposed suboptimal algorithms.

\end{abstract}

\renewcommand{\baselinestretch}{0.94}
\large\normalsize

\section{Introduction}
\label{sect1}

 Orthogonal frequency
division multiplexing (OFDM) is  a promising air interface to fulfill the growing demands for a high spectral efficiency,  due to its flexibility in resource allocation and high resistance  against channel delay spread. Nevertheless, in energy limited wireless networks,  the lifetime of communication nodes remains the bottleneck in guaranteeing quality of service (QoS) due to
 the constrained energy supply. Recently, energy harvesting technology has received considerable interest from both
industry and academia. In particular, it has become a viable  solution for prolonging the lifetime of networks since it provides a perpetual energy source and self-sustainability to systems \cite{JR:Mag_green}\nocite{CN:energy-harvesting,CN:WIPT_fundamental,CN:Shannon_meets_tesla,CN:MIMO_WIPT}-\cite{JR:WIP_receiver}.

Traditionally,  tides, solar, and wind are the major natural renewable energy sources for energy harvesting.  Unfortunately, the availability  of these energy sources is limited by climate or location which may be problematic in indoor environments. On the other hand, harvesting energy from ambient  radio signals in radio frequency (RF) introduces a new paradigm of energy management. More importantly, wireless
energy harvesting provides the possibility for simultaneous wireless information and power transfer \cite{CN:WIPT_fundamental}-\nocite{CN:Shannon_meets_tesla,CN:MIMO_WIPT}\cite{JR:WIP_receiver}. Nevertheless, this combination poses many new challenges for the design of resource allocation algorithms and receivers. In \cite{CN:WIPT_fundamental} and \cite{CN:Shannon_meets_tesla}, the
optimal tradeoff between power and information transfer was studied for different system settings. However,  the receivers in \cite{CN:WIPT_fundamental} and \cite{CN:Shannon_meets_tesla} are assumed to be able to decode information and extract power from the same received signal, which cannot be achieved in practice yet.
 Consequently,  a power splitting receiver was proposed in \cite{CN:MIMO_WIPT} and  \cite{JR:WIP_receiver} for facilitating simultaneous  information decoding and  energy harvesting. In particular,  the authors in  \cite{JR:WIP_receiver} studied the resource allocation algorithm design for power splitting receivers of single carrier systems in  ergodic fading
channels.  Yet, the assumption of channel ergodicity may not be justified in slow fading channels and the results in \cite{JR:WIP_receiver}  may not be applicable to multicarrier systems. Besides, an algorithm for maximizing  the spectral efficiency of a system with power splitting receiver has not been reported in the literature so far.

Motivated by the aforementioned prior works, in this paper,
we first derive an optimal algorithm for maximizing the system spectral efficiency in slow fading channels. Due to the associated  high computational complexity of the optimal algorithm, we propose two suboptimal resource allocation algorithms with low computational complexity which are based on the coordinate ascent method and convex optimization techniques.

 \begin{figure*}\centering\vspace*{-5mm}
\includegraphics[width=5in]{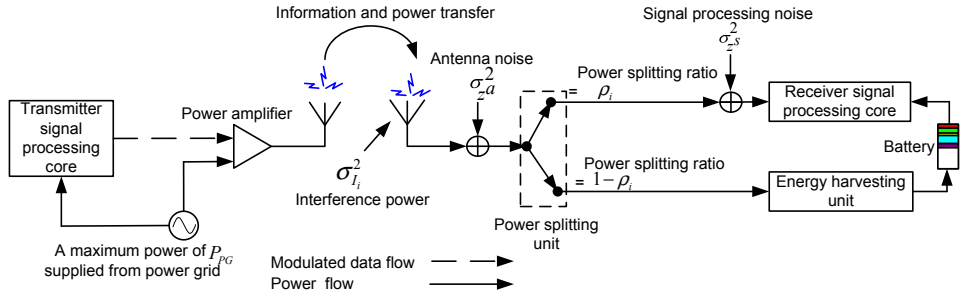}
 \caption{Communication system  model for simultaneous wireless information and power transfer.} \label{fig:system_model}\vspace*{-5mm}
\end{figure*}

\section{System Model}
\label{sect:OFDMA_AF_network_model}
In this section, we present the adopted system model.
\subsection{OFDM Channel Model}
 A point-to-point OFDM system consisting of one transmitter and one receiver is considered.  We assume that the receiver is able to decode information and harvest energy from noise and radio signals (desired signal and interference signal). All transceivers are equipped with a
single antenna, cf.  Figure \ref{fig:system_model}.  There are $n_F$
subcarriers sharing the system
 bandwidth of $\cal B$ Hertz; each subcarrier has a bandwidth of $W={\cal B}/n_F$ Hertz.  The channel impulse
response is assumed to be time invariant (slow fading) and the channel gain is available at the transmitter for resource allocation purpose. In addition,  the receiver is impaired by a co-channel interference signal emitted  by an unintended transmitter. The downlink
received symbol at the receiver
on subcarrier $i\in\{1,\,\ldots,\,n_F\}$ is
given by
\begin{eqnarray}
Y_{i}=\sqrt{P_{i}g l }H_{i}X_{i}+I_i+Z_{i}^s +Z_{i}^a,
\end{eqnarray}
where $X_{i}$, $P_{i}$, and ${H}_{i}$  are the transmitted
symbol,  the transmitted power, and the multipath fading coefficient on subcarrier $i$, respectively.
 $l$ and $g$ denote the
path loss attenuation and shadowing between the transmitter and receiver, respectively. $Z_{i}^s$ is the signal processing noise on subcarrier $i$ with zero mean and variance $\sigma_{z^s}^2$. $Z_{i}^s$ is caused by quantization errors. $Z_{i}^a$ is the antenna noise on subcarrier $i$ which is modeled as additive white Gaussian noise (AWGN) with zero mean and variance $\sigma_{z^a}^2$. $I_i$ is the received interference signal on subcarrier $i$ with zero mean and variance $\sigma_{I_i}^2$.

\subsection{Hybrid  Energy Harvesting and Information Receiver}
\label{sect:receiver}
In practice, the signal used for information decoding cannot be reused for  harvesting energy due to hardware limitations \cite{JR:WIP_receiver}. As a result,  we follow a similar approach as in \cite{JR:WIP_receiver} and focus on a  power splitting receiver to facilitate the concurrent information decoding and energy harvesting. In particular, the receiver splits the received signal into two  power streams carrying proportions of $\rho_i$ and $1-\rho_i$ of the total received signal power before any active analog/digial signal processing is performed, cf. Figure \ref{fig:system_model}. Consequently, the two streams carrying a fraction of $\rho_i$ and $1-\rho_i$ of the total received signal power are used for decoding the information embedded in the signal and  energy harvesting, respectively\footnote{For the sake of presentation, we assume that the power splitting ratio can be different across different subcarriers at the moment. The implementation constraints on the power splitting  will be taken into account when we introduce the problem formulation for resource allocation algorithm design.  }.

\section{Resource Allocation}\label{sect:forumlation}
In this section, we first introduce the system capacity and then formulate the corresponding resource allocation algorithm design as optimization problem.

\subsection{Instantaneous Channel Capacity}
\label{subsect:Instaneous_Mutual_information}
The channel capacity (maximum spectral efficiency) between the transmitter  and the receiver on
subcarrier $i$ with channel bandwidth $W$ is given by
\begin{eqnarray}\label{eqn:cap}
C_{i}&=&W\log_2\Big(1+P_i\Gamma_{i}\Big)\,\,\,\,
\mbox{with}\,\,\\  \label{eqn:SINR}
\Gamma_{i}&=&\frac{\rho_il
g\abs{H_i}^2}{\rho_i(\sigma_{z^a}^2+\sigma_{I_i}^2)+\sigma_{z^s}^2},
\end{eqnarray}
 where $P_i \Gamma_i$ is the received signal-to-interference-plus-noise ratio (SINR) on subcarrier $i$.  Here, the interference and signal processing noise are treated as AWGN for simplifying the algorithm design. The \emph{system capacity} is defined as the aggregate
 number of
 bits  delivered to the receiver over $n_F$ subcarriers and is given by
\begin{eqnarray}
 \label{eqn:avg-sys-goodput} && \hspace*{-5mm} U({\cal P}, {\boldsymbol \rho})=\sum_{i=1}^{n_F} C_{i},
\end{eqnarray}
where ${\cal P}=\{ P_i \ge 0, \forall i\}$ is the power allocation policy and $\boldsymbol{\rho}=\{0\le \rho_i\le 1, \forall i\}$ is the power splitting ratio policy.

\subsection{Optimization Problem Formulation}
\label{sect:cross-Layer_formulation}
The optimal resource allocation policy, $\{{\cal P}^*$, ${\boldsymbol\rho}^*\}$,  can be
obtained by solving the  following optimization problem:

\begin{eqnarray}
\label{eqn:cross-layer}&&\hspace*{10mm} \max_{{\cal P}, \rho
}\,\, U({\cal P}, {\boldsymbol \rho}) \nonumber\\
\notag \hspace*{-5mm}\mbox{s.t.} &&\hspace*{-5mm}\mbox{C1:}
 \sum_{i=1}^{n_F} P_i \abs{H_i}^2g l \eta (1-\rho_i) \ge P_{\min}^{req},\notag\\
&&\hspace*{-5mm}\mbox{C2:}\notag\sum_{i=1}^{n_F}P_i\le P_{\max},\quad \mbox{C3:}\,\, P_C+\sum_{i=1}^{n_F}\varepsilon P_i\le P_{PG},\\ &&\hspace*{-5mm}\notag\mbox{C4: }\rho_i=\rho_j, \forall j\neq i,\,
\,\,\,\mbox{C5:}\,\, P_i\ge 0, \,\, \forall i,
\\&&\hspace*{-5mm} \mbox{C6:}\,\, 0\le\rho_i\le 1, \forall i.
\end{eqnarray}
 Here, $P_{\min}^{req}$ in C1 is the minimum required
 power transfer to the receiver which represents a QoS requirement. $\eta$ denotes the energy harvesting efficiency of the receiver in converting the received
radio signal to electrical energy for storage.
 C2 constrains the maximum transmit power of the transmitter such that it will not exceed $P_{\max}$. In practice, the value of $P_{\max}$ is related to hardware limitations of the transmitter and/or the maximum spectrum mask specified by regulations. Variables $P_C$ and $\varepsilon$ in C3 are two constants account for the circuit power consumption in the transmitter and the inefficiency of the power
amplifier, respectively.  C3 indicates that the maximum  power supply from the power grid is $P_{PG}$, cf. Figure \ref{fig:system_model}, and the total power consumption of the transmitter is controlled to be less than $P_{PG}$. C4 accounts for  the hardware limitations of the power splitting receiver. In particular,  $\rho_{i}$ is required to be identical for all subcarriers. Otherwise,  an analog adaptive passive frequency selective power splitter is required  at the receiver which results in a high system complexity.


\section{Solution of the Optimization Problem} \label{sect:solution}
Problem (\ref{eqn:cross-layer}) belongs to the class  of nonconvex optimization problems. In particular, the power splitting ratio, $\rho_i$, couples with the power allocation variable, $P_i$, in the  SINR  on each subcarrier which makes (\ref{eqn:cap}) a nonconvex function with respect to (w.r.t.) $\rho_i$ and $P_i$. As a result, efficient convex optimization techniques may not be applicable for obtaining the global optimal solution.  In the following, we propose an optimal algorithm and two suboptimal algorithms for system capacity maximization. The optimal resource allocation algorithm comprises a full search for $\rho_i$ and convex optimization techniques. Specifically,
we maximize the system capacity w.r.t. the transmit power for a given fixed $\rho_i$. Then, we repeat the procedure for all possible values of $\rho_i$ and record the corresponding achieved system capacities\footnote{In general, an $n_F$ dimensions full search is required to obtain the optimal power splitting ratio. Yet, the search space can be reduced to a one-dimensional search because of constraint C4. }. At the end,  we select that $\rho_i$ as the optimal power splitting ratio from all the trials which provides the maximum system  capacity. We note that although the optimal resource allocation algorithm achieves the global optimal system performance, it incurs a prohibitively  high computational complexity  to the transmitter which is not desirable for time constrained wireless communication services.


\subsection{Optimal Algorithm}

In this subsection, we solve the power allocation
optimization problem by convex optimization techniques for a given set of $\boldsymbol \rho$.
To this end, we first obtain
the Lagrangian function of (\ref{eqn:cross-layer}):
\begin{eqnarray}\hspace*{-2mm}&&{\cal
L}(  \lambda,\beta,\gamma,{\cal P})\\ \notag
\hspace*{-5mm}&=&\hspace*{-3mm}\sum_{i=1}^{n_F}
C_i\hspace*{-0.5mm}-\hspace*{-0.5mm}\lambda\Big( P_C+\sum_{i=1}^{n_F}\varepsilon P_i- P_{PG}\hspace*{-1mm}\Big)\\
\hspace*{-2mm}&-&\hspace*{-3mm}\beta\Big(\sum_{i=1}^{n_F}P_i- P_{\max}\Big)-\gamma\Big(P_{\min}^{req}-\sum_{i=1}^{n_F} P_i \abs{H_i}^2g l \eta (1-\rho_i)\Big).\notag
\label{eqn:Lagrangian}
\end{eqnarray}
Here, $\gamma,\beta,\lambda\ge0$ are the Lagrange multipliers associated with the transmitter power usage constraints C1, C2, and C3, respectively. The non-negative transmit power constraint in C5  will be captured in the Karush-Kuhn-Tucker (KKT) conditions\footnote{Since the problem is concave w.r.t. $\cal P$ for a given set of ${\boldsymbol \rho}$ and it satisfies Slater's constraint qualification, the KKT conditions provide necessary and sufficient conditions for the optimal transmit power allocation. } when we derive the optimal transmit power.  We note that constraints C4 and C6 do not contribute to the Lagrangian function since they are independent of $P_i$. Yet, these constraints will be considered in the full search over $\rho_i$.

Then, by using the KKT conditions for a fixed set of Lagrange multipliers,  the optimal power allocation
on subcarrier $i$  is obtained as
 \begin{eqnarray}\label{eqn:power1}
\hspace*{-3mm}P_{i}^*\hspace*{-2mm}&=&\hspace*{-2mm}
\Bigg[\frac{W}{\ln(2)(\alpha+\beta-\gamma)}-\frac{\rho_i(\sigma_{z^a}^2+\sigma_{I_i}^2)+\sigma_{z^s}^2}{\rho_il
g\abs{H_i}^2}\Bigg]^+, \,\forall i,
\end{eqnarray}
and $\big[x\big]^+=\max\{0,x\}$. It can be observed from  (\ref{eqn:power1}) that the power allocation solution is in the form of water-filling. Since $\rho_i$ is fixed in this optimization framework, the influence of $\rho_i$ is treated as part of the channel gain. Lagrange multiplier $\gamma$
forces the transmitter to allocate more power for transmission to fulfill the minimum power transfer requirement $P_{\min}^{req}$. The optimal Lagrange multipliers can be easily found via the gradient method or off-the-shelf numerical solvers \cite{book:non_linear_programming,book:convex}.

In the following, we focus on the design of two suboptimal power allocation algorithms with  low computational complexity compared to the optimal power allocation algorithm.

\begin{table}[t]\caption{Iterative Resource Allocation Algorithm.}\label{table:algorithm1}
\vspace*{-4mm}
\begin{algorithm} [H]                    
\renewcommand\thealgorithm{}
\caption{Suboptimal Resource Allocation 1}          
\label{alg1}                           
\begin{algorithmic} [1]
\normalsize           
\STATE Initialization: $N_{\max}=$ the maximum number of iterations  and  $\Delta=$ the
maximum tolerance
 \STATE Set  iteration index $n=0$ and initial resource allocation policy $\{{\cal P}_n, {\boldsymbol{ \rho}}_n\}$

\REPEAT [Iteration]
\STATE For
a  given set of ${\boldsymbol{ \rho}}_n$, obtain an intermediate power allocation ${{\cal P}_n'}$ from (\ref{eqn:power1})
\STATE For
a  given set of ${{\cal P}_n'}$, obtain an intermediate power splitting ratio ${\boldsymbol{ \rho}_n'}$ from (\ref{eqn:rho-subopt1})
\IF {$|U({{\cal P}_n'},{\boldsymbol{ \rho}_n'})-U({\cal P}_n,{\boldsymbol{ \rho}}_n)|<\Delta$} \STATE  $\mbox{Convergence}=\,$\TRUE \RETURN
$\{{\cal P^*,\mbox{\boldmath$\rho$}^*}\}=\{{\cal P}', {\boldsymbol \rho}_n'\}$  \ELSE \STATE
Set $\{{\cal P}_n, {\boldsymbol{ \rho}}_n\}=\{{\cal P}_n', {\boldsymbol{ \rho}}_n'\}$ and $n=n+1$ \STATE  Convergence $=$ \FALSE
 \ENDIF
 \UNTIL{Convergence $=$ \TRUE $\,$or $n=N_{\max}$}

\end{algorithmic}
\end{algorithm}
\vspace*{-1.0cm}
\end{table}
\subsection{Suboptimal Algorithm 1}
The problem formulation  in (\ref{eqn:cross-layer}) is concave w.r.t. $\cal P$  or $\boldsymbol \rho$, but not jointly concave w.r.t. both of them. As a result, an iterative coordinate ascent method is proposed to obtain a locally optimal solution \cite{book:non_linear_programming} of (\ref{eqn:cross-layer}) and the algorithm is summarized in Table \ref{table:algorithm1}. ${\cal P}_n$ and ${\boldsymbol \rho}_n$ denote the power allocation and power splitting policy  in the $n$th iteration, respectively.  The overall algorithm is implemented by a loop which solves two optimization problems iteratively. In each iteration, we execute line 4 in Table  \ref{table:algorithm1}  by  first keeping the power splitting ratio ${\boldsymbol{ \rho}_n}$ fixed and optimizing  ${{\cal P}_n}$ via (\ref{eqn:power1}). Then in line 5, we keep the updated transmit power ${{\cal P}_n}$ fixed and optimize ${\boldsymbol{ \rho}_n}$. The procedure iterates between line 4 and line 5 until the algorithm converges or the maximum number of iterations has been reached. We note that the convergence to a local optimal solution is guaranteed \cite{book:non_linear_programming}.

For solving the  power splitting ratio ${\boldsymbol{ \rho}}_n$ with a fixed ${{\cal P}}_n$ in line 5, we apply the KKT conditions for (\ref{eqn:cross-layer}) which yields

\begin{eqnarray}\label{eqn:rho-subopt1}
\hspace*{-3mm}\rho_{i}^*\hspace*{-2mm}&=&\Bigg[\frac{\Theta_i/2/ (\sigma_{z^a}^2+\sigma_{I_i}^2)/ (\gamma+\sum_{j\ne i}\sum_i \zeta_{i,j})}{ \sqrt{\ln(2)}\, \left((\sigma_{z^a}^2+\sigma_{I_i}^2) + \abs{H_i}^2lgP_i \right)}\notag\\
 &&- \frac{(\sigma_{z^a}^2+\sigma_{I_i}^2)\, \sigma_{z^s}^2 + \frac{\abs{H_i}^2lgP_i  \sigma_{z^s}^2}{2}}{(\sigma_{z^a}^2+\sigma_{I_i}^2)\, \left((\sigma_{z^a}^2+\sigma_{I_i}^2) + \abs{H_i}^2lgP_i \right)}\Bigg]_0^1
\end{eqnarray}
where $\Theta_i=\sqrt{\abs{H_i}^2lgP_i  \sigma_{z^s}^2\, (\gamma+\sum_{j\ne i}\sum_i \zeta_{i,j}) \Phi_i}$, $\Phi_i=4W {(\sigma_{z^a}^2+\sigma_{I_i}^2)}^2 + 4W\abs{H_i}^2lgP_i (\sigma_{z^a}^2+\sigma_{I_i}^2) + \abs{H_i}^2lgP_i  \sigma_{z^s}^2\, (\gamma+\sum_{j\ne i}\sum_i \zeta_{i,j}) \ln(2)$, and $\zeta_{i,j}$ is a two-dimensional  Lagrange multiplier chosen to satisfy the consensus constraint C4.
 Operator $\big[x\big]^c_d$ is defined as  $\big[x\big]^c_d=c,\ \mbox{if}\ x>c,\big[x\big]^c_d=x,\mbox{ if}\, d\le x\le c,\big[x\big]^c_d  =d,\ \mbox{if}\ d>x$, respectively. It can be verified that  $\rho_{i}^*$ is a monotonic increasing function of $P_i$, i.e., $\frac{\partial \rho_{i}^* }{\partial P_i} >0$. As a result, we expect that when $P_i$ is large enough, $\rho_{i}^*$ will have a value close to 1. We note that the proposed suboptimal algorithm  has a polynomial time complexity due to the convexity of the problem formulation w.r.t. the individual optimization variables.

\subsection{Suboptimal Algorithm 2}
In this section, we propose a suboptimal resource allocation algorithm which is asymptotically optimal in the high SINR regime.  For facilitating the design of an efficient resource allocation algorithm, we augment the optimization variable space by replacing $\rho_i$ with two auxiliary  variables, i.e., $\rho_i^I$ and $\rho_i^E$. Specifically, $\rho_i^I$ and $\rho_i^E$ are associated with the power splitting ratios of the power streams for information decoding and energy harvesting, respectively. Note that $\rho_i^I+\rho_i^E=1$ has to be satisfied which indicates that the power splitting unit does not introduce any extra power gain to the received signal. In addition,  we rewrite and approximate  the channel capacity between the transmitter and the receiver on subcarrier $i$ as
\begin{eqnarray}\label{eqn:approx_cap}
C_{i}&=&W\log_2\Big(1+\frac{P_i\rho_i^Il
g\abs{H_i}^2}{\rho_i^I(\sigma_{z^a}^2+\sigma_{I_i}^2)+\sigma_{z^s}^2}\Big)\\
&\approx& W\log_2\Big(\frac{P_i\rho_i^Il
g\abs{H_i}^2}{\rho_i^I(\sigma_{z^a}^2+\sigma_{I_i}^2)+\sigma_{z^s}^2}\Big)\notag
\end{eqnarray}
in high SINR, i.e., $\log_2(1+x)\approx\log_2(x), x\gg 1$. As a result, the objective function is now jointly concave w.r.t. $\rho_i^I$ and $P_i$ since the two eigenvalues of the Hessian matrix of $C_i$, $\frac{-1}{P_i^2\log(2)}$ and $-\frac{(\sigma_{z^s}^2)^2 + \rho_i^I\left(2\, \sigma_{z^a}^2 + 2\, \sigma_{I_i}^2\right)\, \sigma_{z^s}^2}{({\rho_i^I})^2 \log(2) {\left(\sigma_{z^s}^2 + \rho_i^I\, \sigma_{z^a}^2 + \rho_i^I \sigma_{I_i}^2\right)}^2}$, are  non-positive. Furthermore, we rewrite constraints C1, C4, and C6 as
\begin{eqnarray}
&&\mbox{C1}:
 \sum_{i=1}^{n_F} P_i \abs{H_i}^2g l \eta \ge \frac{P_{\min}^{req}}{\rho_1^E},\\
&&\mbox{C4}: \rho_1^E=\rho_r^E, \forall r=\{2,\ldots,n_F\} , \quad\mbox{and}\\
&&\mbox{C6}:
 \rho_i^I+\rho_i^E=1,\forall i,
\end{eqnarray}
respectively. Finally, we impose a non-negative value constraint on the auxiliary variables, $\mbox{C7: } \rho_i^I,\rho_i^E\ge0,\forall i$. Hence, the constraints C1--C7 span a convex feasible solution set. Therefore, the problem with the approximated channel capacity (\ref{eqn:approx_cap}) is jointly concave w.r.t. the optimization variables and traditional convex optimization techniques can be used for obtaining the  solution.   The Lagrangian of the transformed problem with the approximated objective function is given by
\begin{eqnarray}\hspace*{-8mm}&&{\cal
L}(  \lambda,\beta,\gamma,{\boldsymbol
\mu},{\boldsymbol
\phi},{\cal P},{\boldsymbol
\rho})\notag\\ \notag
\hspace*{-8mm}&=&\hspace*{-3mm}\sum_{i=1}^{n_F}
C_i\hspace*{-0.5mm}-\hspace*{-0.5mm}\lambda\Big( P_C+\sum_{i=1}^{n_F}\varepsilon P_i- P_{PG}\hspace*{-1mm}\Big)-\sum_{r=2}^{n_F}\phi_r(\rho_1^E-\rho_r^E)\\
\hspace*{-5mm}&-&\hspace*{-3mm}\beta\Big(\sum_{i=1}^{n_F}P_i- P_{\max}\Big)-\gamma\Big(\frac{P_{\min}^{req}}{ \rho_1^E}-\sum_{i=1}^{n_F} P_i \abs{H_i}^2g l \eta\Big)\notag\\
\hspace*{-5mm}&-&\hspace*{-3mm}\sum_{i=1}^{n_F}\mu_i(\rho_i^I+\rho^E_i-1),
\label{eqn:Lagrangian2}
\end{eqnarray}
where $\boldsymbol{\mu}$ and $\boldsymbol{\phi}$ are the Lagrange multiplier vectors with elements $\mu_{i}$ and $\phi_j$ which are associated with constraints C6 and C4, respectively. Therefore, the transmit power and the power splitting factor  can be obtained as

 \begin{eqnarray}\label{eqn:power-subopt2}
\hspace*{-3mm}P_{i}^*\hspace*{-2mm}&=&\hspace*{-2mm}
\Bigg[\frac{W}{\ln(2)(\alpha+\beta-\gamma)}\Bigg]^+, \,\forall i,\\
\hspace*{-3mm}\rho_{i}^{I*}\hspace*{-2mm}&=&\Bigg[\frac{\sqrt{\mu_i\, \sigma_{z^s}^2\, \left(4W(\sigma_{z^a}^2+\sigma_{I_i}^2) + \mu_i\, \sigma_{z^s}^2\, \ln\!\left(2\right)\right)}}{2\, \mu_i\,(\sigma_{z^a}^2+\sigma_{I_i}^2)\, \sqrt{\ln\!\left(2\right)}}
\\
&&- \frac{\sigma_{z^s}^2}{2\,(\sigma_{z^a}^2+\sigma_{I_i}^2)}\Bigg]^+, \forall i,\notag\\
\hspace*{-3mm}\rho_{r}^{E*}\hspace*{-2mm}&=&\Big[\mu_r+\phi_r\Big]^+,\,\,\forall r\{2,\ldots,n_F\}, \quad\mbox{and}\\
\hspace*{-3mm}\rho_{1}^{E*}\hspace*{-2mm}&=&\Big[\frac{P_{\min}^{req}\gamma}{\mu_1+\sum_{j=1}^{n_F}\phi_j}\Big]^+.
\end{eqnarray}
 The power allocation solution in (\ref{eqn:power-subopt2}) suggests that  equal power allocation across different subcarriers is optimal in the high SINR regime. Moreover, $P_{i}^*$ is decoupled from $\rho_{i}^{I*}$ and, in contrast to suboptimal algorithm 1, iteration between $P_{i}^*$ and $\rho_{i}^{I*}$ is not required for obtaining the solution.
Since the considered problem with the approximated objective function is jointly concave w.r.t. the optimization variables, the problem can be solved efficiently by finding the optimal Lagrange multipliers with numerical solvers \cite{book:convex}.

\begin{table}[t]\vspace*{-2mm}\caption{System parameters}\label{tab:feedback} \centering
\begin{tabular}{|l|l|}\hline
 \mbox{Receiver distance} & \mbox{10 meters}  \\

    \hline
 \mbox{Multipath fading distribution} & \mbox{Rician  fading (Rician factor  6 dB) }  \\

    \hline
\mbox{Carrier center frequency} & 470 MHz  \\
  \hline
\mbox{Number of subcarriers $n_F$} &  128  \\
    \hline
    \mbox{Total bandwidth } &  20 MHz   \\

    \hline
    \mbox{Signal processing noise } $\sigma_{z^s}^2$ &  \mbox{$-35$ dBm}   \\
    \hline
        \mbox{Antenna noise} $\sigma_{z^a}^2$ &  \mbox{$-115$ dBm}   \\
    \hline
\mbox{Channel path loss model} &  \mbox{TGn path loss model}    \\
    \hline
\mbox{Lognormal shadowing $g$} &  \mbox{1}   \\
    \hline
    \mbox{Circuit power consumption $P_C$} &  40 \mbox{dBm}   \\
        \hline
    \mbox{Max. power grid supply $P_G$} &  50 \mbox{dBm}   \\
    \hline
            \mbox{Power amplifier power efficiency } & $1/{\varepsilon}=0.16$  \\

                \hline
                            \mbox{Min. required power transfer} $P_{\min}^{req}$ & $0$ dBm  \\

                \hline
                                            \mbox{Energy harvesting efficiency} $\eta$ & $0.8$  \\

                \hline
\end{tabular}\vspace*{-4mm}
\end{table}

\section{Simulations}
\label{sect:result-discussion} In this section, we evaluate the
 performance of the proposed power allocation algorithms using simulations. The simulation parameters can be found in
Table \ref{tab:feedback}. For the optimal resource allocation algorithm, we use 1000 equally spaced intervals  for  quantizing  the range of $\rho_i$ for facilitating  the full search. Besides, the maximum number of iterations $N_{\max}$ in suboptimal algorithm 1 is set to 5. Note that if the transmitter
is unable to guarantee the  minimum required power transfer $P_{\min}^{req}$, we
set the system capacity for that channel realization to zero to account for the corresponding failure. For the sake of illustration, we define the interference-to-signal processing noise ratio (INR) as $\frac{ \sigma_{I_{i}}^2}{\sigma_{z^s}^2}$.

 \begin{figure}[t]
\centering\vspace*{-4mm}
\includegraphics[width=3.5 in]{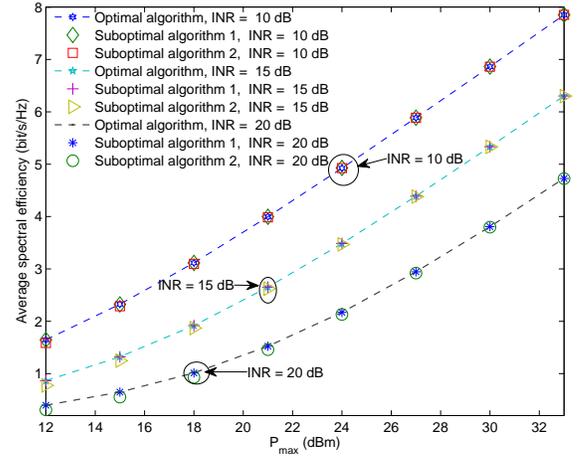}\vspace*{-4mm}
 \caption{Average system spectral efficiency (bit/s/Hz) versus
maximum transmit power allowance, $P_{\max}$, for different  levels of INR, $\frac{\sigma_{I_{i}}^2}{\sigma_{z^s}^2}$.} \label{fig:CAP_PT}
\end{figure}

\subsection{Average System Spectral Efficiency}
Figure \ref{fig:CAP_PT} plots the  average system spectral efficiency (bit/s/Hz) of the three proposed algorithms versus the
maximum transmit power allowance, $P_{\max}$, for different INRs, $\frac{\sigma_{I_{i}}^2}{\sigma_{z^s}^2}$.
It can be observed that the spectral efficiency of all algorithms increases with increasing $P_{\max}$.
This is attributed to the fact that the transmitters in the algorithms  radiate all the power available at the transmitter at every time instant whenever it is possible. Besides,  an increasing amount of INR impairs the spectral efficiency of the system, despite the fact that part of the energy can be harvested by the receiver for satisfying the minimum required power transfer. On the other hand, suboptimal algorithm 1 performs very close to the optimal algorithm in all considered scenarios due to the iterative optimization. Besides, as expected, suboptimal algorithm 2 gives an excellent performance when $P_{\max}$ is large since it is asymptotically optimal in the high SINR regime. Yet, the performance of suboptimal algorithm 2 is less appealing in the low SINR regime, e.g. for $\mbox{INR}=20$ dB and $P_{\max}<18$ dBm, compared to the other two algorithms, due to the approximation of the objective function, cf. (\ref{eqn:power-subopt2}).

\subsection{Average Harvested Power and Power Splitting Ratio}
\begin{figure}[t]
\centering\vspace*{-4mm}
\includegraphics[width=3.5 in]{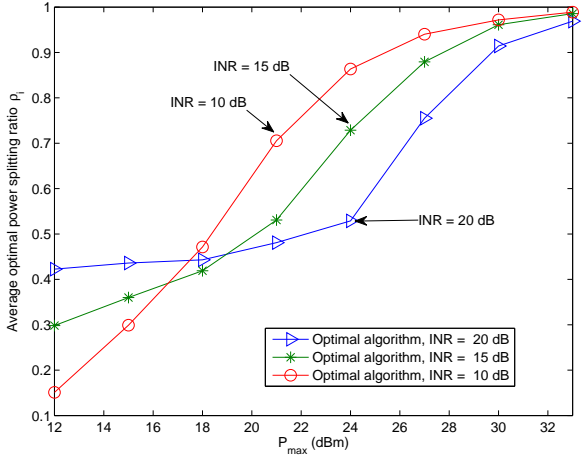}\vspace*{-4mm}
 \caption{Average optimal power splitting ratio, $\rho_i$, of the optimal algorithm versus the
 maximum transmit power allowance, $P_{\max}$,  for different  levels of INR, $\frac{\sigma_{I_{i}}^2}{\sigma_{z^s}^2}$.}  \label{fig:rho1}
\vspace*{-4mm}
\end{figure}

\begin{figure}[t]
\centering\vspace*{-4mm}
\includegraphics[width=3.5 in]{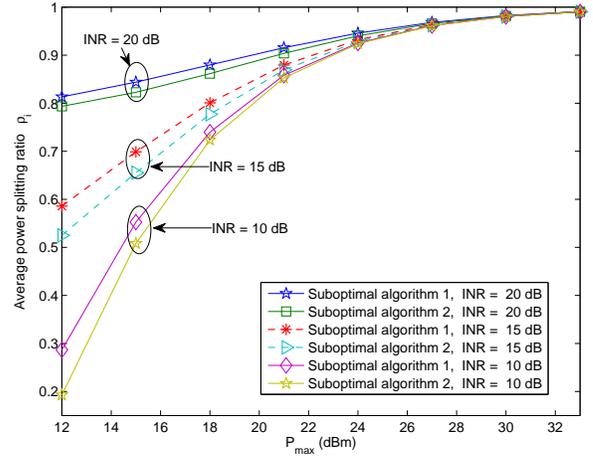}\vspace*{-4mm}
 \caption{Average  power splitting ratio, $\rho_i$, of the two suboptimal algorithms versus the
 maximum transmit power allowance, $P_{\max}$,  for different  levels of INR, $\frac{\sigma_{I_{i}}^2}{\sigma_{z^s}^2}$. } \label{fig:rho2}
\end{figure}

\begin{figure}[t]
 \centering\vspace*{-4mm}
\includegraphics[width=3.5 in]{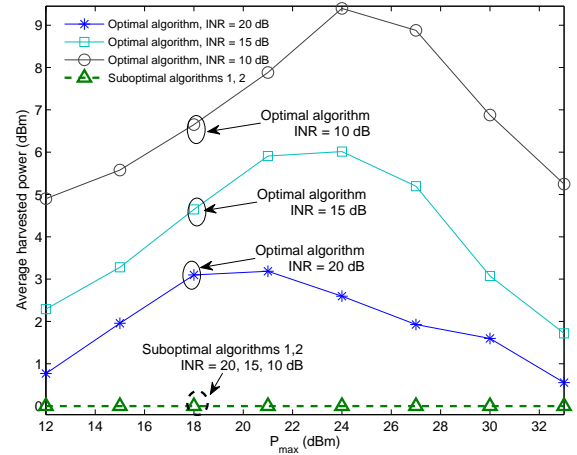}\vspace*{-4mm}
\caption{Average harvested power (dBm) versus the
 maximum transmit power allowance, $P_{\max}$,  for different  levels of INR, $\frac{\sigma_{I_{i}}^2}{\sigma_{z^s}^2}$.} \label{fig:pt_pt}\vspace*{-4mm}
\end{figure}

Figures \ref{fig:rho1} and \ref{fig:rho2} show, respectively, the average power splitting ratio, $\rho_i$, of the optimal algorithm and the two suboptimal algorithms versus
maximum allowed transmit power, $P_{\max}$, for different levels of INR, $\frac{\sigma_{I_{i}}^2}{\sigma_{z^s}^2}$. For the optimal algorithm, it can be observed in Figure \ref{fig:rho1} that in all considered scenarios, the values of $\rho_i$ increase w.r.t. to the maximum transmit power allowance $P_{\max}$.  Although splitting more power for information decoding could also possibly increase the associated interference power, $\rho_i(\sigma_{z^a}^2+\sigma_{I_i}^2)$, the power gain due to an increasing $\rho_i$ in the signal strength  of the desired signal, $\rho_i P_i \abs{H_i}^2$, is able to counteract the performance impairment since the desired signal strength dominates the total received power. Besides, the slope of the curves depends heavily on the values of INRs and $P_{\max}$.  Indeed, the role of  $\rho_i$ is to balance the spectral efficiency caused by $\rho_i P_i \abs{H_i}^2$ and the degradation caused by $\rho_i(\sigma_{z^a}^2+\sigma_{I_i}^2)$ which results in a non-trivial trade-off between INR, $P_{\max},$ and $\rho_i$, cf. (\ref{eqn:cap}), (\ref{eqn:SINR}). For instance, for a fixed small value of  $P_{\max}$ (e.g. $P_{\max}<18$ dBm), the average optimal value of $\rho_i$ is higher for INR$=$ 20 dB than for INR $=$ 10 dB. Yet, for a fixed large value of  $P_{\max}$ (e.g. $P_{\max}>18$ dBm), the average optimal value of $\rho_i$ is higher for INR$=$ 10 dB than for INR $=$ 20 dB. The detailed trade-off between these variables will be investigated in  future work. On the other hand, Figure \ref{fig:rho2} shows that the values of $\rho_i$ for the two suboptimal algorithms increase monotonically w.r.t. $P_{\max}$. In particular, both suboptimal algorithms have a higher preference for splitting more power for information decoding in the high INR regime.  This is because the suboptimal algorithms only split enough power to satisfy  the minimum required power transfer constraint C1 with equality (see Figure \ref{fig:pt_pt}) and allocate the remaining power for information decoding.

Figure  \ref{fig:pt_pt} reveals that, for the optimal algorithm, the average harvested power is a bell-shape function of the maximum transmit power allowance $P_{\max}$. Besides, the amount of harvested power is always larger than the minimum required power transfer    $P_{\min}^{req}$. On the contrary, for the two suboptimal algorithms, the receivers only harvests just enough power for satisfying the minimum required power $P_{\min}^{req}$ which limits the power gain $P_{i}\rho_i$ in the channel capacity. Indeed, the two suboptimal algorithms both achieve a lower average system capacity and a lower average harvested power compared to the optimal algorithm. This is because the two suboptimal algorithms are not able to fully exploit   the spectral efficiency gain achieved by $\rho_i P_i \abs{H_i}^2$ and to reduce the degradation caused by $\rho_i(\sigma_{z^a}^2+\sigma_{I_i}^2)$.


\section{Conclusions}\label{sect:conclusion}
In this paper, we formulated the power allocation
algorithm design for simultaneous  wireless information and power transfer in OFDM systems as a nonconvex optimization problem.
 The problem formulation took into account the minimum required power transfer for  system capacity maximization.
 The problem was solved by a one-dimensional full search and convex optimization techniques which incurred a high complexity at the transmitter.  Hence, two low-complexity suboptimal iterative algorithms were proposed to find a good compromise between  computational complexity and system performance. Simulation results illustrated that the proposed suboptimal algorithms achieved a close-to-optimal system capacity.

\bibliographystyle{IEEEtran}
\bibliography{OFDMA-AF}

\begin{thebibliography}{1}
\providecommand{\url}[1]{#1}
\csname url@samestyle\endcsname
\providecommand{\newblock}{\relax}
\providecommand{\bibinfo}[2]{#2}
\providecommand{\BIBentrySTDinterwordspacing}{\spaceskip=0pt\relax}
\providecommand{\BIBentryALTinterwordstretchfactor}{4}
\providecommand{\BIBentryALTinterwordspacing}{\spaceskip=\fontdimen2\font plus
\BIBentryALTinterwordstretchfactor\fontdimen3\font minus
  \fontdimen4\font\relax}
\providecommand{\BIBforeignlanguage}[2]{{%
\expandafter\ifx\csname l@#1\endcsname\relax
\typeout{** WARNING: IEEEtran.bst: No hyphenation pattern has been}%
\typeout{** loaded for the language `#1'. Using the pattern for}%
\typeout{** the default language instead.}%
\else
\language=\csname l@#1\endcsname
\fi
#2}}
\providecommand{\BIBdecl}{\relax}
\BIBdecl

\bibitem{JR:Mag_green}
T.~Chen, Y.~Yang, H.~Zhang, H.~Kim, and K.~Horneman, ``{Network Energy Saving
  Technologies for Green Wireless Access Networks},'' \emph{IEEE Wireless
  Commun.}, vol.~18, pp. 30--38, Oct. 2011.

\bibitem{CN:energy-harvesting}
D.~W.~K. Ng and R.~Schober, ``{Energy-Efficient Power Allocation for M2M
  Communications with Energy Harvesting Transmitter},'' in \emph{Proc. IEEE
  Global Telecommun. Conf.}, Dec. 2012.

\bibitem{CN:WIPT_fundamental}
L.~Varshney, ``{Transporting Information and Energy Simultaneously},'' in
  \emph{Proc. IEEE Intern. Sympos. on Inf. Theory}, Jul. 2008, pp. 1612 --1616.

\bibitem{CN:Shannon_meets_tesla}
P.~Grover and A.~Sahai, ``{Shannon Meets Tesla: Wireless Information and Power
  Transfer},'' in \emph{Proc. IEEE Intern. Sympos. on Inf. Theory}, 2010, pp.
  2363 --2367.

\bibitem{CN:MIMO_WIPT}
R.~Zhang and C.~K. Ho, ``{MIMO Broadcasting for Simultaneous Wireless
  Information and Power Transfer},'' in \emph{Proc. IEEE Global Telecommun.
  Conf.}, Dec. 2011, pp. 1 --5.

\bibitem{JR:WIP_receiver}
\BIBentryALTinterwordspacing
X.~Zhou, R.~Zhang, and C.~K. Ho, ``{Wireless Information and Power Transfer:
  Architecture Design and Rate-Energy Tradeoff},'' \emph{submitted for possible
  journal publication}, 2012. [Online]. Available:
  \url{http://arxiv.org/abs/1205.0618}
\BIBentrySTDinterwordspacing

\bibitem{book:non_linear_programming}
D.~P. Bertsekas, \emph{{Nonlinear Programming}}, 2nd~ed.\hskip 1em plus 0.5em
  minus 0.4em\relax {Athena Scientific}, 1999.

\bibitem{book:convex}
S.~Boyd and L.~Vandenberghe, \emph{{Convex Optimization}}.\hskip 1em plus 0.5em
  minus 0.4em\relax {Cambridge University Press}, 2004.

\end{thebibliography}

\end{document}